\documentclass[12pt,twocolumn]{aastex631}
\usepackage{amsmath}
\usepackage{epsf}
\usepackage{color}
\usepackage{graphicx}
\usepackage{color}
\usepackage{enumitem}
\usepackage{mathtools}
\usepackage{tabularx}
\usepackage{longtable}
\usepackage{float}

\begin{document}
\title{Cold Quasar Investigation: Comparing Star Formation Rates to Black Hole Growth}
\author{Sasha Mintz}
\affiliation{Department of Physics, Virginia Polytechnic Institute and State University, 850 West Campus Drive,
Blacksburg, VA 24061}
\author{Brandon Coleman}
\author{Allison Kirkpatrick}
\affiliation{Department of Physics \& Astronomy, University of Kansas, 1251 Wescoe Hall Dr., Lawrence, KS 66045}

\begin{abstract}
Cold quasars are a rare population of luminous, unobscured quasars associated with host galaxies that have a high star formation rate. We aimed to study the host galaxies of 64 of these cold quasars in order to probe how the supermassive black holes and host galaxies were co-evolving. We compiled data from the XMM-XXL survey and cross-matched with the VHS, WISE, and HerMES surveys to obtain multiwavelength photometry spanning the X-ray to the infrared and including optical spectroscopy. From the data, we calculated the supermassive black hole’s mass using broad emission from the MgII and Hbeta lines. We compared this with the stellar mass of the entire galaxy and find that the black holes are significantly more massive than would be predicted by local relations, indicating that the majority of black hole growth precedes the bulk of the the stellar mass formation. In addition to this, we created a spectral energy distribution for each galaxy to calculate the star formation rate. We compared the star formation rate with the black hole accretion rate and find that the stellar mass is rapidly increasing at a relative rate faster than the black hole growth, supporting the picture where the black hole grows first.
\end{abstract}


\section{Introduction}

Likely every massive galaxy hosts a supermassive black hole (SMBH) at its center.
The  most active period 
of supermassive black hole accretion and feedback 
is predicted by theoretical models to occur after the galaxy has 
formed most of its stars \citep{Hopkins, Dekel, byrne2023}. In these models, star formation is triggered by major mergers of two massive gas and dust-rich 
galaxies, followed by sudden rapid accretion onto the 
supermassive black hole \citep{Sanders, Petric}. 
The end of this phase 
is signified by the active galactic nucleus (AGN) launching extremely powerful winds that push away any obscuring 
dust \citep{Glikman, Murray}. 
This merger scenario is predicted to be the leading cause of extremely luminous quasars \citep{Hopkins}.

The co-evolution of supermassive black holes and their host galaxies is a vibrant area of astronomical research, with many open questions. As discussed in the simple merger scenario above, feedback emanating from the AGN 
can launch large scale winds that propagate through the host galaxy \citep{Pontzen, Fabian}. These large scale winds may be responsible for quenching star formation (``negative feedback'') through the removal of gas, if the mass loading factor is significant \citep{carniani2017,chen2022,revalski2022} On the other hand, theoretical and observational studies also show that these winds may trigger star formation (``positive feedback'') by compressing molecular clouds \citep{Ciotti}. 

 AGN and quasars (we define a ``quasar'' as an AGN with $L_X>10^{44}$erg/s) whose host galaxies are gas-rich and still forming stars at a rapid pace are also observed in the literature \citep{xie2021,scholtz2021,molina2023, Perna, Kirkpatrick2019}. \citet{Kirkpatrick} argued that the most extreme of these quasars should form a new category, called ``Cold Quasars''. In these galaxies, there is a 
significant presence of cold gas in addition to an extremely high star formation rate (SFR$>200\,M_\odot$/yr). These galaxies represent a rare population where unobscured quasars are found to reside in starbursting galaxies \citep{Coleman, Cooke}.

Directly measuring the impact of postive or negative feedback through winds is only possible with spectroscopy, but co-evolution can be inferred by looking at global relations in large populations. The most well-known
relation is the tight correlation that is observed between the mass of the supermassive black hole ($M_\bullet$) and the mass of stars ($M_\ast$) in the host galaxy \citep{Kormendy}. This correlation 
persists over several orders of magnitude in the local universe, and is observed at higher redshift, although with some evolution of the normalization \citep{somerville2009,sarria2010,schulze2011}. This tight correlation has been used to infer a regulatory process between black hole growth and star formation \citep{silk2012,navarro2020}, although proving causation observationally within galaxies remains elusive.

In unosbscured quasars, identifying a correlation between the black hole and host galaxy masses and mass growth rates (through SFR or black hole accretion rate, BHAR) suffers from the difficulty of separating AGN from host emission.
Unobscured quasars are extremely luminous, and in the mid-infrared and optical, the quasars outshine their hosts, so that typical SFR tracers (e.g., H$\alpha$) cannot be used \citep{Baldwin, Kewley, Kauffmann}. Thus, the best wavelength to study the host galaxy SFR is in the far infrared, as unobscured AGN within star forming galaxies do not heat the dust substantially beyond $\lambda$ $>$ 100 $\mu$m \citep{mullaney2011,kirkpatrick2012,Kirkpatrick,sokol2023}.

In this paper, we trace the growth of the black holes in Cold Quasars through $M_\bullet$ and BHAR. In Section 2, we discuss selecting Cold Quasars from the XMM-XXL survey. In Section 3, we detail how measurements were made. We compare the black hole properties with host $M_\ast$ and SFR. We present our conclusions in Section 4. In this work, we assume a standard flat cosmology with $H_0=70\,$Mpc/km/s, $\Omega_M=0.3$, and $\Lambda=0.7$.


\section{Data}
A significant section of the sky must be surveyed in order to locate 
rare, star-forming quasars. The XXL survey was chosen as the primary data source for this project due to its large area and public multiwavelength data set. The northern XXL field spans $\sim18\,$deg$^2$ and was surveyed with the X-ray Multi-Mirror Mission (XMM-Newton) down to a depth of $F_{\rm, 0.5-10\,keV}=10^{-15}\,$erg/s/cm$^2$ \citep{XXL}. The public XXL catalog contains more than 8000 X-ray sources \citep{menzel2016,liu2016}. The survey was designed with the specific goal of finding high luminosity X-ray emission, which are characteristic of AGNs. The XXL data release also includes spectroscopic redshifts from the Sloan Digital Sky Survey \citep[SDSS;][]{menzel2016}.

In order to enhance the dataset, we cross-matched between the XXL survey and several additional astronomical surveys, including the 
Wide-field Infrared Survey Explorer (WISE), Herschel Multi-tiered Extragalactic Survey (HerMES), and the VISTA Hemisphere Survey (VHS) \citep{WISE, HerMES, VHS}. Full details of the cross-matching can be found in Coleman et al.\,2023 (in prep). We briefly summarize. X-ray sources are point-like, which can make identifying the correct association challenging. The maximum likelihood estimator (MLE) method considers three factors to determine correct counterparts to the
X-ray sources: 1) the density of background
sources in the field as a function of magnitude, 2) the
density of sources around the X-ray
source per unit magnitude after subtracting the background
distribution, and 3) the positional errors associated with each
catalog \citep{ananna2017}. We found that the entirety of the cold quasar candidate subsample examined in this work had unambiguous matches. 


{\it Herschel}/SPIRE is critical for identifying cold quasars. The HERMES survey covers 9.3 deg$^2$ of the XXL field \citep{HerMES}. The $5\sigma$ depth of the SPIRE 250\,$\mu$m image is 25.8\,mJy.

To identify potential cold quasars, we began with 8,445 XXL X-ray sources. There were a total of 3,042 sources that had a unique optical counterpart from SDSS, and 2,525 (83\%) had a spectroscopic redshift attached to them. Out of that, there were 1,772 sources that had multiwavelength counterparts in VHS and WISE. 

We identified a total of 293 galaxies to be cold quasar candidates. These were determined by the following criteria set forth in \cite{Kirkpatrick}:
\begin{enumerate}
    \item X-ray luminosity greater than $10^{44}$ erg/s
    \item Detected in all three {\it Herschel}/SPIRE bands
    \item Absolute B-band magnitude ($M_B$) $<$ -23
\end{enumerate}
To transform between the SDSS filters and the B band, we used the equation
\begin{equation}
m_B=m_g+0.17(m_u-m_g)+0.11
\end{equation}
\citep{jester}.
We refer to this sample of 293 galaxies as cold quasar candidates. We explore their possible starbursting nature in Section \ref{sec:ms}.


\begin{figure}
    \centering
    \includegraphics[width=1.1\linewidth]{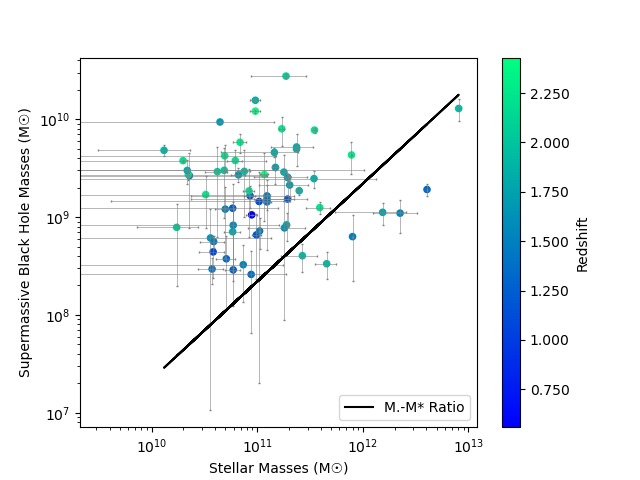}
    \caption{The comparison of supermassive black hole mass, $M_\bullet$, to stellar mass, $M_\ast$, for  our final sample of 64 cold quasar candidates. Points have been shaded by redshift, corresponding to the colorbar on the right. The black linear line represents the $M_{\rm bulge}-M_\bullet$ relationship measured from local galaxies \citep{Kormendy}. Nearly all of the cold quasar candidates lie above the local relation, regardless of redshift. This indicates that the stellar mass must increase more, proportionally, for the galaxies to reach the local relation by $z=0$.} 
    \label{fig:mbh}
\end{figure}

\section{Results and Discussion}
\subsection{\texorpdfstring{$M_\bullet$}{M\_bullet} Calculation}
In order to determine the masses of the supermassive black holes within the cold quasar candidates, we fit the optical broad lines with {\tt PyQSOFit} \citep{PyQSOFit}. {\tt PyQSOFit} fits each optical spectrum with a model of continuum emission due to both an accretion disk and a possible stellar population \citep{yip2004a,yip2004b}, Fe{\sc ii} templates \citep{boroson1992,vestergaard2001}, a Balmer continuum \citep{dietrich2002}, and a series of broad lines. The program returns values such as emission line strength, full width at half maximum (FWHM), and luminosity values at varying wavelengths (the luminosities at 3000\,\AA\ and 5100\,\AA, $L_{3000}$ and $L_{5000}$ are relevant for this work).
We used {\tt PyQSOFit} to measure the luminosity and FWHM of the Mg{\sc ii} and H$\beta$ emission lines. If the spectrum contained neither, due to insufficient wavelength coverage, it was removed from the sample. 
From the original cold quasar candidate sample, we were able to fit Mg{\sc ii} for 245 galaxies and H$\beta$ for seven galaxies.
For the Mg{\sc ii} candidates, we used the following equation from \cite{Kollmeier} 
to calculate the mass of the supermassive black hole:
\begin{equation}[H]
    M_\bullet = 2.04\left(\frac{L_{3000}}{10^{44}\,{\rm erg\,s^{-1}}}\right)^{0.88}\left(\frac{\rm FWHM_{\rm Mg{\sc ii}}}{\rm km\,s^{-1}}\right)^2
\end{equation}
There are multiple calibrations in the literature for calculating masses. We also calculate Mg{\sc ii} masses following Equation 2 in \cite{Trakhtenbrot} and find them to be consistent within the uncertainties.
 
For the H$\beta$ candidates, we calculated the mass with: 
\begin{equation}
    M_\bullet = 8.13\left(\frac{L_{5100}}{10^{44}\,\rm erg\,s^{-1}}\right)^{0.5}\left(\frac{\rm FWHM_{\rm H\beta}}{\rm km\,s^{-1}}\right)^2
\end{equation}
from \cite{Kollmeier} as well. The uncertainties on $M_\bullet$ are dominated by the uncertainties on the FWHM of the broad lines. To estimate the $M_\bullet$ uncertainty, we ran a Monte Carlo simulation where we resampled the FWHM of the broad line, assuming a normal distribution of the associated uncertainty. 
We then recalculated $M_\bullet$ with the resampled FWHM. We ran this simulation 10,000 times. The final uncertainty is the standard deviation of the $M_\bullet$ distribution.
After calculating the uncertainties on each sample, those that had higher uncertainty than the mass itself were removed from the dataset. This left a total of 64 galaxies.

\subsection{\texorpdfstring{The $M_\bullet$-$M_\ast$ Relation for Cold Quasars}{The M\_bullet-M\_ast Relation for Cold Quasars}}
 We calculate $M_\ast$ for the final 64 galaxies using the fitting code {\tt SED3Fit} \citep{berta2013}. {\tt SED3Fit} fits the unattenuated stellar emission, attenuated stellar emission, and dust emission from star formation using an energy balance technique. The code uses stellar libraries from \citet{bruzual2003} and IR-dust libraries from \citet{dacunha2008}.  {\tt SED3Fit} also takes into account the possible contribution from an AGN accretion disk and torus using the library from \citet{feltre2012}. 
 
 We show the $M_\bullet$ versus $M_\ast$ in Figure \ref{fig:mbh} as a function of redshift. To understand how $M_\bullet$ compares to other galaxies, we also overplot the ratio of black hole mass to bulge mass in local galaxies from \citet{Kormendy}. It is notable that the majority of our sample lie above this relation. The highest redshift sources in general host the most massive black holes and lie furthest away from the local relation. The measurement of stellar mass can be uncertain if the AGN dominates the optical portion of the SED. It is notable, however, that most galaxies still lie above the local relation even given the large uncertainties on $M_\ast$. To reach the local relation by $z=0$, the stellar mass in cold quasar candidates needs to dramatically increase. $M_\bullet$, on the other hand, is already on par with some of the most massive black holes in the local universe.

\begin{deluxetable*}{r c c c c c c c c}
\tablecaption{Properties of Cold Quasar Candidates}
\tablehead{\colhead{Galaxy ID} & \colhead{RA} & \colhead{Dec} & \colhead{$z$} & 
\colhead{FWHM} & \colhead{$M_\bullet$} & \colhead{$M_\ast$} & \colhead{SFR} & \colhead{BHAR}\\
\colhead{} & \colhead{(J2000)} & \colhead{(J2000)} & \colhead{} & \colhead{(km/s)} & \colhead{($10^{8}$$M_\odot$)} &
\colhead{($10^{11}$$M_\odot$)} & \colhead{($M_\odot$/yr)} & \colhead{$10^{44}$erg/sec}}
\startdata
        N\_4\_19 & 02:27:40.9 & -3:52:50.40 & 1.931 & 2922.9 & $(4.03 \pm 1.3)$ & $(2.03 \pm 1.0)$ & $(893 \pm 364.8)$ & $(1.70 \pm 0.08)$\\
        N\_9\_19 & 02:27:32.3 & -3:27:36.60 & 1.783 & 2426.4 & $(11.2 \pm 2.8)$ & $(15.4 \pm 10)$ & $(387.3 \pm 162.9)$ & $(3.20 \pm 0.15)$\\
        N\_11\_7 & 02:24:28.8 & -3:15:33.32 & 1.353 & 4915.5 & $(3.75 \pm 2.6)$ & $(0.59 \pm 0.1)$ & $(394.3 \pm 152.3)$ & $(0.66 \pm 0.01)$\\
        N\_11\_22 & 02:24:22.2 & -3:10:52.63 & 1.225 & 3093.3 & $(12.3 \pm 9.6)$ & $(8.00 \pm 1.0)$ & $(252.8 \pm 95.5)$ & $(2.88 \pm 0.13)$ \\
        N\_12\_35 & 02:25:36.5 & -3:28:33.15 & 2.205 & 6744.6 & $(37.8 \pm 11.2)$ & $(4.55 \pm 1.0)$ & $(933.3 \pm 387.6)$ & $(1.72 \pm 0.10)$\\
        N\_12\_40 & 02:25:14.5 & -3:10:59.27 & 2.246 & 4359.7 & $(12.5 \pm 1.7)$ & $(0.62 \pm 0.1)$ & $(477.5 \pm 212.0)$ & $(1.30 \pm 0.08)$\\
        N\_18\_41 & 02:22:34.7 & -4:04:17.69 & 2.043 & 3921.7 & $(7.85 \pm 5.89)$ & $(0.86 \pm 0.1)$ & $(1326.6 \pm 542.5)$ & $(5.06 \pm 0.16)$\\  
        N\_27\_14 & 02:22:54.9 & -4:35:8.03 & 1.221 & 13953.5 & $(15.3 \pm 7.3)$ & $(0.88 \pm 0.1)$ & $(187.7 \pm 70.1)$ & $(1.18 \pm	0.32)$\\
        N\_28\_45 & 02:23:19.0 & -4:46:14.24 & 1.979 & 10996.9 & $(29.3 \pm 19.2)$ & $(0.17 \pm 0.1)$ & $(163.7 \pm 72.1)$ & $(1.06 \pm 0.06)$\\
        N\_30\_35 & 02:25:57.6 & -4:50:05.46 & 2.275 & 6345.5 & $(58.1 \pm 13.2)$ & $(1.94 \pm 1.0)$ & $(382.2 \pm 159.8)$ & $(1.23 \pm 0.14)$\\
        N\_31\_30 & 02:24:9.8 & -4:47:18.15 & 1.819 & 6194.2 & $(25.4 \pm 1.4)$ & $(0.75 \pm 0.1)$ & $(922.6 \pm 374.1)$ & $(1.42 \pm 0.09)$\\
        N\_35\_40 & 02:24:24.6 & -5:04:14.59 & 2.048 & 11867.1 & $(30 \pm 2.03)$ & $(0.83 \pm 0.1)$ & $(518.8 \pm 213.9)$ & $(4.98 \pm 0.23)$ \\
        N\_36\_31 & 02:23:12.4 & -5:06:25.09 & 2.205 & 7747.5 & $(37.7 \pm 1.0)$ & $(0.69 \pm 0.1)$ & $(411.0 \pm 173.7)$ & $(5.47 \pm 0.37)$\\
        N\_37\_12 & 02:19:34.7 & -5:06:28.04 & 2.293 & 9738.6 & $(43.1 \pm 15.4)$ & $(1.94 \pm 1.0)$ & $(588.3 \pm 242.0)$ & $(15.4 \pm 1.3)$\\
        N\_37\_57 & 02:19:08.3 & -4:55:01.41 & 1.843 & 12959.7 & $(45.7 \pm 5.1)$ & $(0.20 \pm 0.1)$ & $(154.1 \pm 66.3)$ & $(4.53 \pm	0.17)$\\
        N\_38\_79 & 02:18:30.5 & -4:56:22.9 & 1.397 & 4501.9 & $(26.5 \pm 18.9)$ & $(7.81 \pm 1.0)$ & $(302.2 \pm 116.7)$ & $(1.31 \pm 0.08)$\\        
        N\_39\_5 & 02:19:26.7 & -4:45:05.17 & 1.467 & 5057.7 & $(56.2 \pm 7.9)$ & $(0.23 \pm 0.1)$ & $(600.0 \pm 233.9)$ & $(2.26 \pm 0.06)$\\
        N\_39\_19 & 02:19:03.5 & -4.:39:35.06 & 1.843 & 11151.0 & $(52.0 \pm 18.4)$ & $(1.45 \pm 1.0)$ & $(432.0 \pm 176.6)$ & $(1.82 \pm 0.07)$ \\
        N\_41\_24 & 02:16:43.7 & -5:22:36.36 & 1.851 & 10073.2 & $(49.6 \pm 2.1)$ & $(2.36 \pm 1.0)$ & $(433.8 \pm 177.4)$ & $(1.74 \pm 0.16)$\\
        N\_42\_6 & 02:18:34.5 & -5:25:52.49 & 1.537 & 5372.5 & $(3.26 \pm 1.9)$ & $(0.39 \pm 0.1)$ & $(617.5 \pm 243.5)$ & $(2.24 \pm 0.10)$ \\
        N\_44\_59 & 02:26:44.2 & -4:07:20.29 & 1.478 & 5953.4 & $(16.4 \pm 4.6)$ & $(2.36 \pm 1.0)$ & $(396.8 \pm 155.3)$ & $(2.11 \pm 0.10)$ \\
        N\_44\_71 & 02:27:23.7 & -4:02:15.01 & 1.822 & 7513.2 & $(7.08 \pm 5.84)$ & $(0.74 \pm 0.1)$ & $(712.0 \pm 289.2)$ & $3.48 \pm	0.16)$\\
        N\_47\_22 & 02:27:29.2 & -4:32:27.86 & 2.261 & 8938.9 & $(79.7 \pm 26.1)$ & $(1.24 \pm 1.0)$ & $(1273.6 \pm 519.0)$ & $(2.88 \pm 0.11)$\\
        N\_47\_34 & 02:28:39.4 & -4:29:12.43 & 1.735 & 11564.7 & $(21.2 \pm 5.3)$ & $(0.22 \pm 0.1)$ & $(203.8 \pm 84.4)$ & $(2.45 \pm 0.16)$ \\
        N\_47\_50 & 02:27:33.9 & -4:25:23.51 & 1.585 & 5383.8 & $(8.25 \pm 6.01)$ & $(0.59 \pm 0.1)$ & $(332.6 \pm 132.7)$ & $(2.32 \pm 0.09)$\\
        N\_48\_27 & 02:23:50.7 & -4:31:58.30 & 1.504 & 3863.1 & $(7.20 \pm 2.48)$ & $(1.71 \pm 1.0)$ & $(189.2 \pm 75.3)$ &  $(4.18 \pm 0.11)$ \\        
        N\_50\_22 & 02:25:25.6 & -4:35:09.65 & 2.165 & 10687.6 & $(27.1 \pm 3.3)$ & $(0.59 \pm 0.1)$ & $(339.0 \pm 142.1)$ & $(1.15 \pm 0.06)$ \\
        N\_51\_9 & 02:26:09.2 & -4:36:50.88 & 1.813 & 12965.0 & $(29.9 \pm 8.0)$ & $(1.05 \pm 0.1)$ & $(277.9 \pm 114.4)$ & $(2.69 \pm 0.08)$  \\        
        N\_54\_30 & 02:21:24.5 & -5:02:05.40 & 1.411 & 4195.8 & $(14.3 \pm 09.7)$ & $(2.68 \pm 1.0)$ & $(210.8 \pm 82.8)$ & $(2.15 \pm 0.11)$\\
        N\_56\_24 & 02:21:58.8 & -5:29:32.38 & 1.353 & 7126.1 & $(19.1 \pm 2.9)$ & $(0.22 \pm 0.1)$ & $(174.5 \pm 68.1)$ & $(4.94 \pm	0.15)$\\
        N\_62\_16 & 02:25:42.3 & -5:29:50.79 & 1.286 & 2559.0 & $(28.9 \pm 1.0)$ & $(1.24 \pm 1.0)$ & $(234.1 \pm 88.9)$ & $(1.78 \pm	0.05$\\        
        N\_63\_24 & 02:24:31.5 & -5:28:18.91 & 2.085 & 7066.5 & $(276 \pm 11.0)$ & $(40.5 \pm 10.0)$ & $(1302.1 \pm 533.3)$ & $(1.61 \pm	0.08)$\\
        N\_66\_41 & 02:23:05.0 & -5:42:50.94 & 1.068 & 3754.7 & $(4.40 \pm 2.04)$ & $(0.59 \pm 0.1)$ & $(151.0 \pm 54.7)$ & $(2.39 \pm  0.10)$\\
        N\_71\_35 & 02:20:33.4 & -6:00:22.16 & 1.695 & 10143.8 & $(93.6 \pm 0.70)$ & $(0.89 \pm 0.1)$ & $(456.1 \pm 187.1)$ & $(5.27 \pm 0.23)$ \\
        N\_75\_4 & 02:17:55.0 & -5:39:21.38 & 1.882 & 5595.1 & $(18.6 \pm 1.8)$ & $(0.44 \pm 0.1)$ & $(247.2 \pm 103.1)$ & $(2.15 \pm 0.08)$ \\
        N\_77\_4 & 02:20:01.6 & -5:22:17.08 & 2.221 & 8760.7 & $(77.3 \pm 4.4)$ & $(0.49 \pm 0.1)$ & $(426.5 \pm 176.8)$ & $(1.70 \pm 0.07)$\\
        N\_77\_15 & 02:19:38.8 & -5:18:05.57 & 2.223 & 8227.1 & $(42.2 \pm 12.4$ & $(2.50 \pm 1.0)$ & $(200.0 \pm 88.5)$ & $(1.08 \pm 0.06)$\\
        N\_78\_16 & 02:15:32.0 & -5:15:19.11 & 1.743 & 6930.9 & $(32.1 \pm 10.1)$ & $(3.49 \pm 1.0)$ & $(148.8  \pm 63.1)$ & $(14.2 \pm	1.2)$\\
        N\_83\_23 & 02:14:19.7 & -4:25:29.83 & 2.174 & 12947.8 & $(18.5 \pm 12.3)$ & $(1.49 \pm 1.0)$ & $(802.6 \pm 335.6)$ & $(6.55 \pm 0.21)$\\
        N\_86\_2 & 02:16:30.8 & -4:20:51.53 & 1.529 & 5484.7 & $(12.0 \pm 8.3)$ & $(0.50 \pm 0.1)$ & $(379.7 \pm 150.0)$ & $(8.01 \pm 0.46)$\\
        N\_90\_4 & 02:17:14.5 & -4:00:20.97 & 1.402 & 4041.9 & $(2.95 \pm 0.87)$ & $(0.37 \pm 0.1)$ & $(235.1 \pm 91.8)$ & $(3.60 \pm 0.12)$\\
        N\_96\_11 & 02:13:24.7 & -3:35:11.44 & 1.141 & 6280.8 & $(14.4 \pm 1.42)$ & $(1.04 \pm 1.0)$ & $(347.3 \pm 127.1)$ & $(3.18 \pm 0.10)$\\
        N\_97\_1 & 02:23:20.6 & -5:41:24.97 & 1.753 & 8445.0 & $(25.5 \pm 2.1)$ & $(1.93 \pm 1.0)$ & $(301.5 \pm 125.5)$ & $(4.36 \pm 0.22)$\\
        N\_99\_3 & 02:14:31.7 & -5:28:00.14 & 1.643 & 4479.0 & $(7.74 \pm 6.9)$ & $(1.81 \pm 1.0)$ & $(272.2 \pm 110.7)$ & $(4.33 \pm 0.27)$ \\
    \label{tab:CQ Table}
\enddata
\end{deluxetable*}

\begin{deluxetable*}{r c c c c c c c c}
\tablecaption{Continuation of Properties of Cold Quasar Candidates}
\tablehead{\colhead{Galaxy ID} & \colhead{RA} & \colhead{Dec} & \colhead{$z$} & 
\colhead{FWHM} & \colhead{$M_\bullet$} & \colhead{$M_\ast$} & \colhead{SFR} & \colhead{BHAR}\\
\colhead{} & \colhead{(J2000)} & \colhead{(J2000)} & \colhead{} & \colhead{(km/s)} & \colhead{($10^{11}$$M_\odot$)} &
\colhead{($10^{11}$$M_\odot$)} & \colhead{($M_\odot$/yr)} & \colhead{erg/s}}

\startdata
        N\_100\_18 & 02:21:32.67 & -03:44:00.8 & 1.605 & 7099.4 & $(15.2 \pm 1.5)$ & $(1.24 \pm 1.0)$ & $(640.9 \pm 256.4)$ & $(14.9 \pm 1.9)$\\
        N\_101\_4 & 02:13:39.1 & -5:42:31.35 & 2.063 & 12645.0 & $(29.1 \pm 23.1)$ & $(3.44 \pm 1.0)$ & $(726.1 \pm 303.0)$ & $(2.11 \pm 0.10)$\\
        N\_101\_11 & 02:14:16.4 & -5:39:44.94 & 1.644 & 6820.4 & $(6.11 \pm 6.0)$ & $(0.361 \pm 0.1)$ & $(1056.1 \pm 423.1)$ & $(3.16 \pm 0.17)$\\
        N\_101\_25 & 02:14:04.1 & -5:33:31.61 & 2.428 & 8855.3 & $(27.0 \pm 18.0)$ & $(1.87 \pm 1.0)$ & $(1129.3 \pm 466.5)$ & $(2.99 \pm 0.14)$\\
        N\_101\_28 & 02:14:37.3 & -5:30:10.65 & 1.854 & 5688.4 & $(24.7 \pm 5.1)$ & $(1.17 \pm 1.0)$ & $(440.4 \pm 181.6)$ & $(3.97 \pm 0.09)$ \\
        N\_102\_8 & 02:28:04.6 & -4:15:59.28 & 2.254 & 8218.5 & $(16.9 \pm 9.2)$ & $(0.13 \pm 0.1)$ & $(459.9 \pm 190.8)$ & $(1.23 \pm 0.06)$\\
        N\_102\_22 & 02:29:01.3 & -4:08:35.41 & 1.524 & 6923.23 & $(11.0 \pm 4.1)$ & $(0.42 \pm 0.1)$ & $(554.1 \pm 218.1)$ & $(9.66 \pm 0.71)$ \\
        N\_102\_36 & 02:28:28.1 & -4:00:44.5 & 1.903 & 4114.0 & $(48.1 \pm 6.3)$ & $(22.5 \pm 10.0)$ & $(240.4 \pm 101.5)$ & $(2.58 \pm 0.18)$\\
        N\_103\_5 & 02:28:49.1 & -4:36:26.98 & 1.935 & 19012.9 & $(129.0 \pm 33.0)$ & $(0.33 \pm 0.10)$ & $(829.4 \pm 339.0)$ & $(1.09 \pm 0.03)$\\
        N\_103\_9 & 02:28:45.5 & -4:33:50.39 & 1.871 & 5224.3 & $(156 \pm 5.0)$ & $(80.8 \pm 10.0)$ & $(748.3 \pm 305.0)$ & $(1.10 \pm 0.10)$\\
        N\_105\_18 & 02:30:00.9 & -4:28:35.40 & 1.486 & 3402.4 & $(6.31 \pm 4.07)$ & $(0.96 \pm 0.10)$ & $(355.6 \pm 140.6)$ & $(8.89 \pm 0.26)$\\
        N\_122\_31 & 02:13:54.9 & -4:05:46.89 & 1.930 & 5462.5 & $(3.33 \pm 1.0)$ & $(0.51 \pm 0.10)$ & $(371.4 \pm 160.5)$ & $(3.53 \pm 0.03)$\\
        N\_131\_26 & 02:13:12.7 & -4:30:46.38 & 1.801 & 4758.9 & $(8.32 \pm 2.64)$ & $(3.91 \pm 1.0)$ & $(514.3 \pm 212.7)$ & $(4.56 \pm 0.21)$\\
        N\_135\_19 & 02:11:29.8 & -4:33:24.00 & 2.430 & 12178.8 & $(121.0 \pm 4)$ & $(1.90 \pm 1.0)$ & $(429.9 \pm 228.2)$ &  $(2.15 \pm 0.07)$\\
        N\_141\_23 & 02:12:47.4 & -5.:06:11.13 & 1.811 & 11782.8 & $(28.8 \pm 14.3)$ & $(0.96 \pm 0.1)$ & $(386.6 \pm 164.0)$ & $(3.30 \pm 0.14)$\\  
        N\_142\_7 & 02:12:16.3 & -5:14:00.43 & 1.230 & 7812.9 & $(6.57 \pm 4.25)$ & $(1.80 \pm 1.0)$ & $(247.7 \pm 94.1)$ & $(1.73 \pm 0.05)$\\
        N\_152\_6 & 02:15:51.8 & -6:17:33.71 & 1.225 & 4869.9 & $(16.5 \pm 4.2)$ & $(0.48 \pm 0.10)$ & $(389.9 \pm 145.8)$ & $(22.8 \pm 7.3)$\\
        N\_152\_47 & 02:15:29.2 & -6:00:46.40 & 1.658 & 7317.9 & $(27.0 \pm 4.4)$ & $(0.98 \pm 0.1)$ & $(239.1 \pm 102.7)$ & $(4.42 \pm 0.07)$ \\
        N\_153\_21 & 02:13:35.3 & -6:01:33.49 & 1.372 & 3787.1 & $(2.59 \pm 1.94)$ & $(0.67 \pm 0.1)$ & $(312.2 \pm 125.2)$ & $(18.2 \pm 3.4)$ \\
    \label{tab:CQ table contd}
\enddata
\tablenotetext{}{FWHM is measured from the MgII or H$\beta$ line using PyQSOFit.}
\end{deluxetable*}

\begin{figure}
    \centering
    \includegraphics[width=1.1\linewidth]{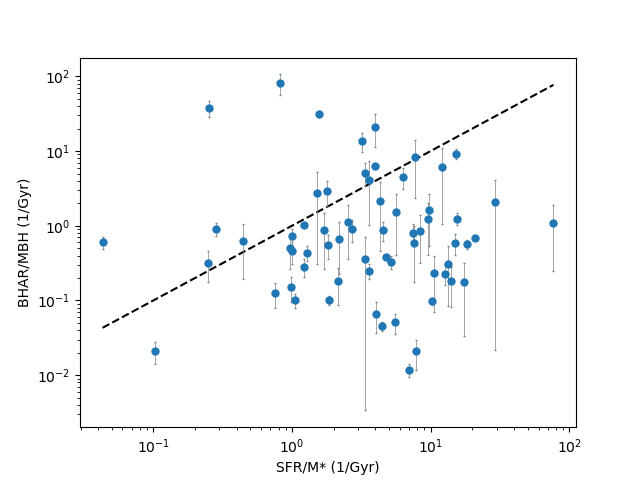}
    \caption{The comparison of normalized black hole accretion rate vs. specific star formation rate for our sample. A one-to-one correlation is plotted as the dashed line. Our cold quasar candidates are proportionally growing their stellar mass much faster than their black hole mass.}
    \label{fig:BHAR}
\end{figure}

\subsection{Measuring Star Formation Rate and Black Hole Accretion Rate}
We calculate the star formation rates (SFRs) for the cold quasar candidates using the 250\,$\mu$m photometry. We opt not to use the SFR output by {\tt SED3Fit}, as {\tt SED3Fit} underpredicts the far-IR emission for many of these galaxies, due to their extreme dust content.
The spectral energy distribution (SED) for cold quasars is dominated by a power-law in the mid-infrared, but the quasar does not contribute to the the heating of the dust beyond $\lambda>50\,\mu$m, even assuming the most generous far-IR AGN heating models \citep{Kirkpatrick}. Therefore, the 250\,$\mu$m emission can be attributed to the star-forming host. Quasars that significantly contribute to the energy budget in the far-IR in very dusty galaxies are heavily obscured, which these quasars are not \citep{sajina2012,ricci2017}. 
We calculate $L_{\rm 250}$ in the rest-frame for each galaxy and then  scale a star-forming template from \citet{kirkpatrick2012} to this luminosity. The star-forming template represents the empirical emission of dusty star-forming galaxies ($L_{\rm IR}>10^{11}\,L_\odot$) at $z\sim1$. We integrate the scaled template from $8-1000\,\mu$m to obtain $L_{\rm IR}$. We convert $L_{\rm IR}$ to a SFR using 

\begin{equation}
    {\rm SFR}\,[M_\odot/{\rm yr}] = 1.59\times 10^{-10} * L_{\rm IR}\,[L_\odot]
\end{equation}

which assumes a Kroupa IMF \citep{murphy2011,kroupa2001}. 
The SFRs for each of these galaxies was extremely high, ranging from $200-1000\,M_\odot$/yr (see Table \ref{tab:CQ Table}), which is similar to the original cold quasar sample in Stripe82X \citep{Kirkpatrick,Coleman}.

We determine the black hole accretion rate (BHAR) by applying a K-correction \citep[Equation 7;][]{KCorrections} to the X-ray luminosity in order to determine the bolometric luminosity, $L_{\rm Bol}$. We then converted $L_{\rm Bol}$ to BHAR assuming a 10\% radiative efficiency.

In Figure \ref{fig:BHAR}, we compare BHAR/$M_\bullet$ to SFR/$M_\ast$ for our sample. We choose to ``normalize'' BHAR and SFR in this manner, rather than directly comparing the two values, because we are probing a range of masses. Larger galaxies naturally have more gas, driving up BHAR and SFR. Any correlation between the two parameters is then likely a product of the galaxy mass.

Interestingly, Figure \ref{fig:BHAR} shows no correlation between black hole growth and stellar growth as parameterized by BHAR/$M_\bullet$ and SFR/$M_\ast$. In most galaxies, the mass of the stellar population is increasing proportionally faster than the mass of the supermassive black hole (indicated by most galaxies lying below the dashed one-to-one correlation line). 

\begin{figure}
    \centering
    \includegraphics[width=1.1\linewidth]{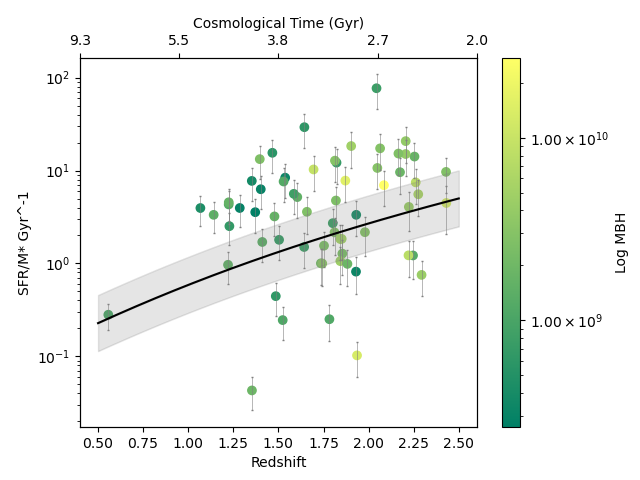}
    \caption{The location of these galaxies on the main sequence relation (black line and grey shaded region) from \citep{elbaz2011}. Galaxies have been shaded by their black hole mass. There is a significant distribution of the galaxies above the main sequence line.}
    \label{fig:main sequence}
\end{figure}

\subsection{Backward Growth?}
\label{sec:ms}
Our sample was selected on the basis of having a detection in all three {\it Herschel}/SPIRE bands, but with the deeper {\it Herschel} observations in the XXL field, many of these candidates may in fact be main sequence galaxies rather than starbursts.

The original sample of Stripe82X cold quasars were found to all have elevated SFRs, placing them in the starburst region \citep{Kirkpatrick,Cooke,Coleman}. Figure \ref{fig:main sequence} investigates the main sequence location of our cold quasar candidates. The main sequence is a tight correlation between SFR and $M_\ast$ observed for the majority of star-forming galaxies \citep{noeske2007,elbaz2007}. We use the main sequence relation parameterized in \citet{elbaz2011} as that was calibrated on {\it Herschel}-detected galaxies out to $z\sim2$. Our cold quasar candidates primarily lie in the starburst regime, above the main sequence line. Interestingly, a few galaxies lie well below the main sequence, despite being detected by {\it Herschel}. According to the criteria in \citep{Coleman}, only the quasars lying above the main sequence are truly cold quasars. 

We shade points according to $M_\bullet$, but the only trend is that the more massive black holes lie at higher redshift. This is driven by the difficulty of detecting smaller black holes with increasing redshift due to survey flux limits.

\begin{figure}
    \centering
    \includegraphics[width=1.1\linewidth]{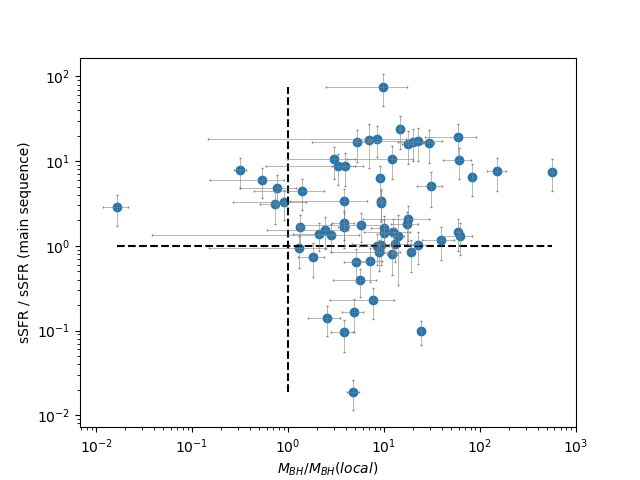}
    \caption{The distance of each galaxy's sSFR from the main sequence versus the distance of each galaxy's supermassive black hole from the local relation \citep{Kormendy}. The main sequence and local relation are shown as the dashed lines. Most of our cold quasar candidates have overmassive black holes given their stellar mass, and they are starbursting, lying in the upper right quadrant. This points to a picture whereby the black hole growth precedes the bulk of the stellar growth.}
    \label{fig:distance}
\end{figure}

Finally, we explore how the high SFRs correlate with high $M_\bullet$ in Figure \ref{fig:distance}, which shows the distance from the main sequence as a function of the distance from the local $M_\bullet-M_\ast$ relation. We have used each galaxy's redshift to predict the sSFR it would have on the main sequence, and we have used each galaxy's $M_\ast$ to predict the $M_\bullet$ it should have if it were on the local relation. 

Most of our sample have an overmassive black hole compared with the local relation, indicating that the stellar population needs to increase in size, in some cases by two orders of magnitude. This increase may be through conversion of the available gas to stars, or it may occur through minor galaxy mergers which increase stellar mass, without significantly affecting the black hole growth \citep{Jahnke, Rodriguez-Gomez}. Similar to Figures \ref{fig:BHAR} and \ref{fig:main sequence}, Figure \ref{fig:distance} does not show any trend between the star forming and black hole properties of a galaxy. Except for a few outliers, all of our cold quasar candidates have overmassive black holes and are starbursting. 

Taken together with Figure \ref{fig:BHAR}, cold quasars seem to be experiencing an evolutionary sequence in which the majority of the supermassive black hole growth predates the majority of the stellar growth. This sort of backward evolution is not common, especially in local galaxies where mergers are driving high SFRs but weaker black hole growth that is heavily obscured \citep{Sanders, Petric}. Merger driven starbursts were the observational impetus for the theoretical models that predict quasar growth occurs after star formation has abated \citep[e.g.]{Hopkins}. Cold quasars, on the other hand, are still strongly star forming, but the black hole growth rate seems to be declining. Over time, each of these galaxy's black holes will cease to accrete further mass, but continue to grow their stellar population, possibly becoming the massive elliptical galaxies we see today. 


\section{Conclusions}
We present a total of 64 galaxies that are to be included within the cold quasar dataset. For each of these 64 galaxies, we measured its supermassive black hole mass using the MgII and H$\beta$ emission from PyQSOFit, using spectra from SDSS. In addition, we measured their stellar mass using SED3Fits. We calculated the star formation rates of each of these galaxies by creating spectral energy distributions and scaling a template of star formation rate to the 250$\mu$m point. Additionally, we measured the black hole accretion rate of each galaxy using K-corrections of their X-ray luminosity, assuming a 10\% radiation efficiency. We conclude the following:

\begin{enumerate}

\item From Figure \ref{fig:mbh}, the supermassive black holes are concluded to be extremely overmassive as compared to local relations. This signifies that the black holes have grown first in these galaxies, and are now completing their evolutionary track.
\item Using Figure \ref{fig:BHAR}, it can be seen that these galaxies' star formation rates are growing proportionally faster than their black holes, implying that their stellar masses need to increase (in some cases double) in order to reach the local relation. The combination of overmassive black holes and increased SFR is indicative of an increase in galaxy growth as well.

\item Figure \ref{fig:main sequence} portrays an interesting conclusion, showing that there are some cold quasars that are starbursting, and a select few are quenching or along the main sequence.
\item Figure \ref{fig:distance} exhibits the conclusion that the majority of the cold quasar candidates have overmassive black holes and are starbursting, with the majority of black hole growth slowing and SFR increasing.

\end{enumerate}

The overview of cold quasars from this study is that they are extremely complex objects. Seeing that there are a select few that are quenching, some that have star formation rates of 1000 solar masses per year, and most contain much larger black holes than predicted exhibits the multi-faceted approach that must be taken in order to begin to understand these quasars. These objects are worth examining in more detail in order to truly understand how these may have come into existence, and how they may be evolving in the distant future.

\vspace{24pt}
\section{Acknowledgements}
SM thanks Meg Urry and Benny Trakhtenbrot for helpful conversations. SM would also like to thank Derek Sikorski for insightful discussions.
We would like the thank the NSF REU Award Number 2149897 for funding SM's research with the University of Kansas.
\linebreak
\linebreak
This research made use of Astropy, a core Python package for Astronomy \citep{astropy}. This project also made use of the Python packages SciPy \citep{SciPy}, Numpy \citep{Numpy}, and Matplotlib \citep{matplotlib}.

\section{Data Availability}
The data used in this paper are available in \cite{menzel2016}. The datasets were derived from sources in the public domain: XMM archive (\url{http://nxsa.esac.esa.int/nxsa-web/#home}), and cross-matched with HerMES (\url{http://hedam.lam.fr/HerMES/index/download}), WISE (\url{https://irsa.ipac.caltech.edu/cgi-bin/Gator/nph-dd}), SDSS (\url{https://www.sdss.org/dr14/data_access/}), and VHS (\url{http://horus.roe.ac.uk:8080/vdfs/VImageList_form.jsp}).

\bibliographystyle{aasjournal}
\bibliography{main}

\begin{thebibliography}{}
\expandafter\ifx\csname natexlab\endcsname\relax\def\natexlab#1{#1}\fi
\providecommand{\url}[1]{\href{#1}{#1}}
\providecommand{\dodoi}[1]{doi:~\href{http://doi.org/#1}{\nolinkurl{#1}}}
\providecommand{\doeprint}[1]{\href{http://ascl.net/#1}{\nolinkurl{http://ascl.net/#1}}}
\providecommand{\doarXiv}[1]{\href{https://arxiv.org/abs/#1}{\nolinkurl{https://arxiv.org/abs/#1}}}

\bibitem[{{Ananna} {et~al.}(2017){Ananna}, {Salvato}, {LaMassa}, {Urry},
  {Cappelluti}, {Cardamone}, {Civano}, {Farrah}, {Gilfanov}, {Glikman},
  {Hamilton}, {Kirkpatrick}, {Lanzuisi}, {Marchesi}, {Merloni}, {Nandra},
  {Natarajan}, {Richards}, \& {Timlin}}]{ananna2017}
{Ananna}, T.~T., {Salvato}, M., {LaMassa}, S., {et~al.} 2017, \apj, 850, 66,
  \dodoi{10.3847/1538-4357/aa937d}

\bibitem[{{Astropy Collaboration} {et~al.}(2013){Astropy Collaboration},
  {Robitaille}, {Tollerud}, {Greenfield}, {Droettboom}, {Bray}, {Aldcroft},
  {Davis}, {Ginsburg}, {Price-Whelan}, {Kerzendorf}, {Conley}, {Crighton},
  {Barbary}, {Muna}, {Ferguson}, {Grollier}, {Parikh}, {Nair}, {Unther},
  {Deil}, {Woillez}, {Conseil}, {Kramer}, {Turner}, {Singer}, {Fox}, {Weaver},
  {Zabalza}, {Edwards}, {Azalee Bostroem}, {Burke}, {Casey}, {Crawford},
  {Dencheva}, {Ely}, {Jenness}, {Labrie}, {Lim}, {Pierfederici}, {Pontzen},
  {Ptak}, {Refsdal}, {Servillat}, \& {Streicher}}]{astropy}
{Astropy Collaboration}, {Robitaille}, T.~P., {Tollerud}, E.~J., {et~al.} 2013,
  \aap, 558, A33, \dodoi{10.1051/0004-6361/201322068}

\bibitem[{Baldwin(2018)}]{Baldwin}
Baldwin, C, e.~a. 2018, Monthly Notices of the Royal Astronomical Society, 473,
  4698

\bibitem[{{Berta} {et~al.}(2013){Berta}, {Lutz}, {Santini}, {Wuyts}, {Rosario},
  {Brisbin}, {Cooray}, {Franceschini}, {Gruppioni}, {Hatziminaoglou}, {Hwang},
  {Le Floc'h}, {Magnelli}, {Nordon}, {Oliver}, {Page}, {Popesso}, {Pozzetti},
  {Pozzi}, {Riguccini}, {Rodighiero}, {Roseboom}, {Scott}, {Symeonidis},
  {Valtchanov}, {Viero}, \& {Wang}}]{berta2013}
{Berta}, S., {Lutz}, D., {Santini}, P., {et~al.} 2013, \aap, 551, A100,
  \dodoi{10.1051/0004-6361/201220859}

\bibitem[{{Boroson} \& {Green}(1992)}]{boroson1992}
{Boroson}, T.~A., \& {Green}, R.~F. 1992, \apjs, 80, 109,
  \dodoi{10.1086/191661}

\bibitem[{{Bruzual} \& {Charlot}(2003)}]{bruzual2003}
{Bruzual}, G., \& {Charlot}, S. 2003, \mnras, 344, 1000,
  \dodoi{10.1046/j.1365-8711.2003.06897.x}

\bibitem[{{Byrne} {et~al.}(2023){Byrne}, {Faucher-Gigu{\`e}re}, {Stern},
  {Angl{\'e}s-Alc{\'a}zar}, {Wellons}, {Gurvich}, \& {Hopkins}}]{byrne2023}
{Byrne}, L., {Faucher-Gigu{\`e}re}, C.-A., {Stern}, J., {et~al.} 2023, \mnras,
  520, 722, \dodoi{10.1093/mnras/stad171}

\bibitem[{{Carniani} {et~al.}(2017){Carniani}, {Marconi}, {Maiolino},
  {Feruglio}, {Brusa}, {Cresci}, {Cano-D{\'\i}az}, {Cicone}, {Balmaverde},
  {Fiore}, {Ferrara}, {Gallerani}, {La Franca}, {Mainieri}, {Mannucci},
  {Netzer}, {Piconcelli}, {Sani}, {Schneider}, {Shemmer}, \&
  {Testi}}]{carniani2017}
{Carniani}, S., {Marconi}, A., {Maiolino}, R., {et~al.} 2017, \aap, 605, A105,
  \dodoi{10.1051/0004-6361/201730672}

\bibitem[{{Chen} {et~al.}(2022){Chen}, {He}, {Ho}, {Gu}, {Wang}, {Zhuang},
  {Liu}, \& {Wang}}]{chen2022}
{Chen}, Z., {He}, Z., {Ho}, L.~C., {et~al.} 2022, Nature Astronomy, 6, 339,
  \dodoi{10.1038/s41550-021-01561-3}

\bibitem[{Ciotti \& Ostriker(2007)}]{Ciotti}
Ciotti, L., \& Ostriker, J.~P. 2007, The Astronomical Journal, 665, 1038

\bibitem[{{Coleman} {et~al.}(2022){Coleman}, {Kirkpatrick}, {Cooke}, {Glikman},
  {LaMassa}, {Marchesi}, {Peca}, {Treister}, {Auge}, {Urry}, {Sanders},
  {Turner}, \& {Ananna}}]{Coleman}
{Coleman}, B., {Kirkpatrick}, A., {Cooke}, K.~C., {et~al.} 2022, \mnras, 515,
  82, \dodoi{10.1093/mnras/stac1679}

\bibitem[{{Cooke} {et~al.}(2020){Cooke}, {Kirkpatrick}, {Estrada}, {Messias},
  {Peca}, {Cappelluti}, {Ananna}, {Brewster}, {Glikman}, {LaMassa}, {Daisy
  Leung}, {Trump}, {Jane Turner}, \& {Urry}}]{Cooke}
{Cooke}, K.~C., {Kirkpatrick}, A., {Estrada}, M., {et~al.} 2020, \apj, 903,
  106, \dodoi{10.3847/1538-4357/abb94a}

\bibitem[{{da Cunha} {et~al.}(2008){da Cunha}, {Charlot}, \&
  {Elbaz}}]{dacunha2008}
{da Cunha}, E., {Charlot}, S., \& {Elbaz}, D. 2008, \mnras, 388, 1595,
  \dodoi{10.1111/j.1365-2966.2008.13535.x}

\bibitem[{Dekel(2014)}]{Dekel}
Dekel, A. 2014, Monthly Notices of the Royal Astronomical Society, 444, 2071

\bibitem[{{Dietrich} {et~al.}(2002){Dietrich}, {Hamann}, {Shields},
  {Constantin}, {Vestergaard}, {Chaffee}, {Foltz}, \&
  {Junkkarinen}}]{dietrich2002}
{Dietrich}, M., {Hamann}, F., {Shields}, J.~C., {et~al.} 2002, \apj, 581, 912,
  \dodoi{10.1086/344410}

\bibitem[{{Elbaz} {et~al.}(2007){Elbaz}, {Daddi}, {Le Borgne}, {Dickinson},
  {Alexander}, {Chary}, {Starck}, {Brandt}, {Kitzbichler}, {MacDonald},
  {Nonino}, {Popesso}, {Stern}, \& {Vanzella}}]{elbaz2007}
{Elbaz}, D., {Daddi}, E., {Le Borgne}, D., {et~al.} 2007, \aap, 468, 33,
  \dodoi{10.1051/0004-6361:20077525}

\bibitem[{{Elbaz} {et~al.}(2011){Elbaz}, {Dickinson}, {Hwang},
  {D{\'\i}az-Santos}, {Magdis}, {Magnelli}, {Le Borgne}, {Galliano},
  {Pannella}, {Chanial}, {Armus}, {Charmandaris}, {Daddi}, {Aussel}, {Popesso},
  {Kartaltepe}, {Altieri}, {Valtchanov}, {Coia}, {Dannerbauer}, {Dasyra},
  {Leiton}, {Mazzarella}, {Alexander}, {Buat}, {Burgarella}, {Chary}, {Gilli},
  {Ivison}, {Juneau}, {Le Floc'h}, {Lutz}, {Morrison}, {Mullaney}, {Murphy},
  {Pope}, {Scott}, {Brodwin}, {Calzetti}, {Cesarsky}, {Charlot}, {Dole},
  {Eisenhardt}, {Ferguson}, {F{\"o}rster Schreiber}, {Frayer}, {Giavalisco},
  {Huynh}, {Koekemoer}, {Papovich}, {Reddy}, {Surace}, {Teplitz}, {Yun}, \&
  {Wilson}}]{elbaz2011}
{Elbaz}, D., {Dickinson}, M., {Hwang}, H.~S., {et~al.} 2011, \aap, 533, A119,
  \dodoi{10.1051/0004-6361/201117239}

\bibitem[{{Fabian}(2012)}]{Fabian}
{Fabian}, A.~C. 2012, \araa, 50, 455,
  \dodoi{10.1146/annurev-astro-081811-125521}

\bibitem[{{Feltre} {et~al.}(2012){Feltre}, {Hatziminaoglou}, {Fritz}, \&
  {Franceschini}}]{feltre2012}
{Feltre}, A., {Hatziminaoglou}, E., {Fritz}, J., \& {Franceschini}, A. 2012,
  \mnras, 426, 120, \dodoi{10.1111/j.1365-2966.2012.21695.x}

\bibitem[{{Glikman} {et~al.}(2015){Glikman}, {Simmons}, {Mailly}, {Schawinski},
  {Urry}, \& {Lacy}}]{Glikman}
{Glikman}, E., {Simmons}, B., {Mailly}, M., {et~al.} 2015, \apj, 806, 218,
  \dodoi{10.1088/0004-637X/806/2/218}

\bibitem[{{Guo} {et~al.}(2018){Guo}, {Shen}, \& {Wang}}]{PyQSOFit}
{Guo}, H., {Shen}, Y., \& {Wang}, S. 2018, {PyQSOFit: Python code to fit the
  spectrum of quasars}, Astrophysics Source Code Library, record ascl:1809.008

\bibitem[{{Hopkins}(2007)}]{Hopkins}
{Hopkins}, A.~M. 2007, in Astronomical Society of the Pacific Conference
  Series, Vol. 380, Deepest Astronomical Surveys, ed. J.~{Afonso}, H.~C.
  {Ferguson}, B.~{Mobasher}, \& R.~{Norris}, 423,
  \dodoi{10.48550/arXiv.astro-ph/0611283}

\bibitem[{{Hunter}(2007)}]{matplotlib}
{Hunter}, J.~D. 2007, Computing in Science and Engineering, 9, 90,
  \dodoi{10.1109/MCSE.2007.55}

\bibitem[{{Jahnke} \& {Macci{\`o}}(2011)}]{Jahnke}
{Jahnke}, K., \& {Macci{\`o}}, A.~V. 2011, \apj, 734, 92,
  \dodoi{10.1088/0004-637X/734/2/92}

\bibitem[{{Jester} {et~al.}(2005){Jester}, {Schneider}, {Richards}, {Green},
  {Schmidt}, {Hall}, {Strauss}, {Vanden Berk}, {Stoughton}, {Gunn},
  {Brinkmann}, {Kent}, {Smith}, {Tucker}, \& {Yanny}}]{jester}
{Jester}, S., {Schneider}, D.~P., {Richards}, G.~T., {et~al.} 2005, \aj, 130,
  873, \dodoi{10.1086/432466}

\bibitem[{{Kauffmann} {et~al.}(2003){Kauffmann}, {Heckman}, {Tremonti},
  {Brinchmann}, {Charlot}, {White}, {Ridgway}, {Brinkmann}, {Fukugita}, {Hall},
  {Ivezi{\'c}}, {Richards}, \& {Schneider}}]{Kauffmann}
{Kauffmann}, G., {Heckman}, T.~M., {Tremonti}, C., {et~al.} 2003, \mnras, 346,
  1055, \dodoi{10.1111/j.1365-2966.2003.07154.x}

\bibitem[{Kewley(1999)}]{Kewley}
Kewley, L, e.~a. 1999, Astrophysics and Space Science, 266, 131

\bibitem[{{Kirkpatrick} {et~al.}(2019){Kirkpatrick}, {Sharon}, {Keller}, \&
  {Pope}}]{Kirkpatrick2019}
{Kirkpatrick}, A., {Sharon}, C., {Keller}, E., \& {Pope}, A. 2019, \apj, 879,
  41, \dodoi{10.3847/1538-4357/ab223a}

\bibitem[{{Kirkpatrick} {et~al.}(2012){Kirkpatrick}, {Pope}, {Alexander},
  {Charmandaris}, {Daddi}, {Dickinson}, {Elbaz}, {Gabor}, {Hwang}, {Ivison},
  {Mullaney}, {Pannella}, {Scott}, {Altieri}, {Aussel}, {Bournaud}, {Buat},
  {Coia}, {Dannerbauer}, {Dasyra}, {Kartaltepe}, {Leiton}, {Lin}, {Magdis},
  {Magnelli}, {Morrison}, {Popesso}, \& {Valtchanov}}]{kirkpatrick2012}
{Kirkpatrick}, A., {Pope}, A., {Alexander}, D.~M., {et~al.} 2012, \apj, 759,
  139, \dodoi{10.1088/0004-637X/759/2/139}

\bibitem[{Kirkpatrick(2020)}]{Kirkpatrick}
Kirkpatrick, A. e.~a. 2020, The Astrophysical Journal, 900

\bibitem[{Kollmeier(2006)}]{Kollmeier}
Kollmeier, J.~A. 2006, The Astrophysical Journal, 648, 128

\bibitem[{{Kormendy} \& {Ho}(2013)}]{Kormendy}
{Kormendy}, J., \& {Ho}, L.~C. 2013, arXiv e-prints, arXiv:1308.6483,
  \dodoi{10.48550/arXiv.1308.6483}

\bibitem[{{Kroupa}(2001)}]{kroupa2001}
{Kroupa}, P. 2001, \mnras, 322, 231, \dodoi{10.1046/j.1365-8711.2001.04022.x}

\bibitem[{{Liu} {et~al.}(2016){Liu}, {Merloni}, {Georgakakis}, {Menzel},
  {Buchner}, {Nandra}, {Salvato}, {Shen}, {Brusa}, \& {Streblyanska}}]{liu2016}
{Liu}, Z., {Merloni}, A., {Georgakakis}, A., {et~al.} 2016, \mnras, 459, 1602,
  \dodoi{10.1093/mnras/stw753}

\bibitem[{Lusso(2012)}]{KCorrections}
Lusso, E. e.~a. 2012, Monthly Notices of the Royal Astronomical Society, 425,
  623

\bibitem[{{Mart{\'\i}n-Navarro} {et~al.}(2020){Mart{\'\i}n-Navarro},
  {Burchett}, \& {Mezcua}}]{navarro2020}
{Mart{\'\i}n-Navarro}, I., {Burchett}, J.~N., \& {Mezcua}, M. 2020, \mnras,
  491, 1311, \dodoi{10.1093/mnras/stz3073}

\bibitem[{{McMahon} {et~al.}(2021){McMahon}, {Banerji}, {Gonzalez}, {Koposov},
  {Bejar}, {Lodieu}, {Rebolo}, \& {VHS Collaboration}}]{VHS}
{McMahon}, R.~G., {Banerji}, M., {Gonzalez}, E., {et~al.} 2021, VizieR Online
  Data Catalog, II/367

\bibitem[{{Menzel} {et~al.}(2016){Menzel}, {Merloni}, {Georgakakis}, {Salvato},
  {Aubourg}, {Brandt}, {Brusa}, {Buchner}, {Dwelly}, {Nandra}, {P{\^a}ris},
  {Petitjean}, \& {Schwope}}]{menzel2016}
{Menzel}, M.~L., {Merloni}, A., {Georgakakis}, A., {et~al.} 2016, \mnras, 457,
  110, \dodoi{10.1093/mnras/stv2749}

\bibitem[{{Molina} {et~al.}(2023){Molina}, {Ho}, {Wang}, {Shangguan}, {Bauer},
  \& {Treister}}]{molina2023}
{Molina}, J., {Ho}, L.~C., {Wang}, R., {et~al.} 2023, \apj, 944, 30,
  \dodoi{10.3847/1538-4357/acaa9b}

\bibitem[{{Mullaney} {et~al.}(2011){Mullaney}, {Alexander}, {Goulding}, \&
  {Hickox}}]{mullaney2011}
{Mullaney}, J.~R., {Alexander}, D.~M., {Goulding}, A.~D., \& {Hickox}, R.~C.
  2011, \mnras, 414, 1082, \dodoi{10.1111/j.1365-2966.2011.18448.x}

\bibitem[{{Murphy} {et~al.}(2011){Murphy}, {Condon}, {Schinnerer}, {Kennicutt},
  {Calzetti}, {Armus}, {Helou}, {Turner}, {Aniano}, {Beir{\~a}o}, {Bolatto},
  {Brandl}, {Croxall}, {Dale}, {Donovan Meyer}, {Draine}, {Engelbracht},
  {Hunt}, {Hao}, {Koda}, {Roussel}, {Skibba}, \& {Smith}}]{murphy2011}
{Murphy}, E.~J., {Condon}, J.~J., {Schinnerer}, E., {et~al.} 2011, \apj, 737,
  67, \dodoi{10.1088/0004-637X/737/2/67}

\bibitem[{{Murray} {et~al.}(2005){Murray}, {Quataert}, \& {Thompson}}]{Murray}
{Murray}, N., {Quataert}, E., \& {Thompson}, T.~A. 2005, \apj, 618, 569,
  \dodoi{10.1086/426067}

\bibitem[{{Noeske} {et~al.}(2007){Noeske}, {Weiner}, {Faber}, {Papovich},
  {Koo}, {Somerville}, {Bundy}, {Conselice}, {Newman}, {Schiminovich}, {Le
  Floc'h}, {Coil}, {Rieke}, {Lotz}, {Primack}, {Barmby}, {Cooper}, {Davis},
  {Ellis}, {Fazio}, {Guhathakurta}, {Huang}, {Kassin}, {Martin}, {Phillips},
  {Rich}, {Small}, {Willmer}, \& {Wilson}}]{noeske2007}
{Noeske}, K.~G., {Weiner}, B.~J., {Faber}, S.~M., {et~al.} 2007, \apjl, 660,
  L43, \dodoi{10.1086/517926}

\bibitem[{{Oliver} {et~al.}(2012){Oliver}, {Bock}, {Altieri}, {Amblard},
  {Arumugam}, {Aussel}, {Babbedge}, {Beelen}, {B{\'e}thermin}, {Blain},
  {Boselli}, {Bridge}, {Brisbin}, {Buat}, {Burgarella},
  {Castro-Rodr{\'\i}guez}, {Cava}, {Chanial}, {Cirasuolo}, {Clements},
  {Conley}, {Conversi}, {Cooray}, {Dowell}, {Dubois}, {Dwek}, {Dye}, {Eales},
  {Elbaz}, {Farrah}, {Feltre}, {Ferrero}, {Fiolet}, {Fox}, {Franceschini},
  {Gear}, {Giovannoli}, {Glenn}, {Gong}, {Gonz{\'a}lez Solares}, {Griffin},
  {Halpern}, {Harwit}, {Hatziminaoglou}, {Heinis}, {Hurley}, {Hwang}, {Hyde},
  {Ibar}, {Ilbert}, {Isaak}, {Ivison}, {Lagache}, {Le Floc'h}, {Levenson},
  {Faro}, {Lu}, {Madden}, {Maffei}, {Magdis}, {Mainetti}, {Marchetti},
  {Marsden}, {Marshall}, {Mortier}, {Nguyen}, {O'Halloran}, {Omont}, {Page},
  {Panuzzo}, {Papageorgiou}, {Patel}, {Pearson}, {P{\'e}rez-Fournon}, {Pohlen},
  {Rawlings}, {Raymond}, {Rigopoulou}, {Riguccini}, {Rizzo}, {Rodighiero},
  {Roseboom}, {Rowan-Robinson}, {S{\'a}nchez Portal}, {Schulz}, {Scott},
  {Seymour}, {Shupe}, {Smith}, {Stevens}, {Symeonidis}, {Trichas}, {Tugwell},
  {Vaccari}, {Valtchanov}, {Vieira}, {Viero}, {Vigroux}, {Wang}, {Ward},
  {Wardlow}, {Wright}, {Xu}, \& {Zemcov}}]{HerMES}
{Oliver}, S.~J., {Bock}, J., {Altieri}, B., {et~al.} 2012, \mnras, 424, 1614,
  \dodoi{10.1111/j.1365-2966.2012.20912.x}

\bibitem[{{Perna} {et~al.}(2018){Perna}, {Sargent}, {Brusa}, {Daddi},
  {Feruglio}, {Cresci}, {Lanzuisi}, {Lusso}, {Comastri}, {Coogan}, {D'Amato},
  {Gilli}, {Piconcelli}, \& {Vignali}}]{Perna}
{Perna}, M., {Sargent}, M.~T., {Brusa}, M., {et~al.} 2018, \aap, 619, A90,
  \dodoi{10.1051/0004-6361/201833040}

\bibitem[{{Petric} {et~al.}(2011){Petric}, {Armus}, {Howell}, {Chan},
  {Mazzarella}, {Evans}, {Surace}, {Sanders}, {Appleton}, {Charmandaris},
  {D{\'\i}az-Santos}, {Frayer}, {Haan}, {Inami}, {Iwasawa}, {Kim}, {Madore},
  {Marshall}, {Spoon}, {Stierwalt}, {Sturm}, {U}, {Vavilkin}, \&
  {Veilleux}}]{Petric}
{Petric}, A.~O., {Armus}, L., {Howell}, J., {et~al.} 2011, \apj, 730, 28,
  \dodoi{10.1088/0004-637X/730/1/28}

\bibitem[{Pierre(2016)}]{XXL}
Pierre, M, e.~a. 2016, A\&A, 592, 16

\bibitem[{Pontzen \& Governato(2012)}]{Pontzen}
Pontzen, A., \& Governato, F. 2012, Monthly Notices of the Royal Astronomical
  Society, 421, 3464

\bibitem[{{Revalski} {et~al.}(2022){Revalski}, {Crenshaw}, {Rafelski},
  {Kraemer}, {Polack}, {Falc{\~a}o}, {Fischer}, {Meena}, {Martinez}, {Schmitt},
  {Collins}, \& {Falcone}}]{revalski2022}
{Revalski}, M., {Crenshaw}, D.~M., {Rafelski}, M., {et~al.} 2022, \apj, 930,
  14, \dodoi{10.3847/1538-4357/ac5f3d}

\bibitem[{{Ricci} {et~al.}(2017){Ricci}, {Assef}, {Stern}, {Nikutta},
  {Alexander}, {Asmus}, {Ballantyne}, {Bauer}, {Blain}, {Boggs}, {Boorman},
  {Brandt}, {Brightman}, {Chang}, {Chen}, {Christensen}, {Comastri}, {Craig},
  {D{\'\i}az-Santos}, {Eisenhardt}, {Farrah}, {Gandhi}, {Hailey}, {Harrison},
  {Jun}, {Koss}, {LaMassa}, {Lansbury}, {Markwardt}, {Stalevski}, {Stanley},
  {Treister}, {Tsai}, {Walton}, {Wu}, {Zappacosta}, \& {Zhang}}]{ricci2017}
{Ricci}, C., {Assef}, R.~J., {Stern}, D., {et~al.} 2017, \apj, 835, 105,
  \dodoi{10.3847/1538-4357/835/1/105}

\bibitem[{{Rodriguez-Gomez} {et~al.}(2016){Rodriguez-Gomez}, {Pillepich},
  {Sales}, {Genel}, {Vogelsberger}, {Zhu}, {Wellons}, {Nelson}, {Torrey},
  {Springel}, {Ma}, \& {Hernquist}}]{Rodriguez-Gomez}
{Rodriguez-Gomez}, V., {Pillepich}, A., {Sales}, L.~V., {et~al.} 2016, \mnras,
  458, 2371, \dodoi{10.1093/mnras/stw456}

\bibitem[{{Sajina} {et~al.}(2012){Sajina}, {Yan}, {Fadda}, {Dasyra}, \&
  {Huynh}}]{sajina2012}
{Sajina}, A., {Yan}, L., {Fadda}, D., {Dasyra}, K., \& {Huynh}, M. 2012, \apj,
  757, 13, \dodoi{10.1088/0004-637X/757/1/13}

\bibitem[{{Sanders} \& {Mirabel}(1996)}]{Sanders}
{Sanders}, D.~B., \& {Mirabel}, I.~F. 1996, \araa, 34, 749,
  \dodoi{10.1146/annurev.astro.34.1.749}

\bibitem[{{Sarria} {et~al.}(2010){Sarria}, {Maiolino}, {La Franca}, {Pozzi},
  {Fiore}, {Marconi}, {Vignali}, \& {Comastri}}]{sarria2010}
{Sarria}, J.~E., {Maiolino}, R., {La Franca}, F., {et~al.} 2010, \aap, 522, L3,
  \dodoi{10.1051/0004-6361/201015696}

\bibitem[{{Scholtz} {et~al.}(2021){Scholtz}, {Harrison}, {Rosario},
  {Alexander}, {Knudsen}, {Stanley}, {Chen}, {Kakkad}, {Mainieri}, \&
  {Mullaney}}]{scholtz2021}
{Scholtz}, J., {Harrison}, C.~M., {Rosario}, D.~J., {et~al.} 2021, \mnras, 505,
  5469, \dodoi{10.1093/mnras/stab1631}

\bibitem[{{Schulze} \& {Wisotzki}(2011)}]{schulze2011}
{Schulze}, A., \& {Wisotzki}, L. 2011, \aap, 535, A87,
  \dodoi{10.1051/0004-6361/201117564}

\bibitem[{{Silk} \& {Mamon}(2012)}]{silk2012}
{Silk}, J., \& {Mamon}, G.~A. 2012, Research in Astronomy and Astrophysics, 12,
  917, \dodoi{10.1088/1674-4527/12/8/004}

\bibitem[{{Sokol} {et~al.}(2023){Sokol}, {Yun}, {Pope}, {Kirkpatrick}, \&
  {Cooke}}]{sokol2023}
{Sokol}, A.~D., {Yun}, M., {Pope}, A., {Kirkpatrick}, A., \& {Cooke}, K. 2023,
  \mnras, 521, 818, \dodoi{10.1093/mnras/stad589}

\bibitem[{{Somerville}(2009)}]{somerville2009}
{Somerville}, R.~S. 2009, \mnras, 399, 1988,
  \dodoi{10.1111/j.1365-2966.2009.15325.x}

\bibitem[{{Trakhtenbrot} \& {Netzer}(2012)}]{Trakhtenbrot}
{Trakhtenbrot}, B., \& {Netzer}, H. 2012, \mnras, 427, 3081,
  \dodoi{10.1111/j.1365-2966.2012.22056.x}

\bibitem[{{van der Walt} {et~al.}(2011){van der Walt}, {Colbert}, \&
  {Varoquaux}}]{Numpy}
{van der Walt}, S., {Colbert}, S.~C., \& {Varoquaux}, G. 2011, Computing in
  Science and Engineering, 13, 22, \dodoi{10.1109/MCSE.2011.37}

\bibitem[{{Vestergaard} \& {Wilkes}(2001)}]{vestergaard2001}
{Vestergaard}, M., \& {Wilkes}, B.~J. 2001, \apjs, 134, 1,
  \dodoi{10.1086/320357}

\bibitem[{{Virtanen} {et~al.}(2020){Virtanen}, {Gommers}, {Oliphant},
  {Haberland}, {Reddy}, {Cournapeau}, {Burovski}, {Peterson}, {Weckesser},
  {Bright}, {van der Walt}, {Brett}, {Wilson}, {Millman}, {Mayorov}, {Nelson},
  {Jones}, {Kern}, {Larson}, {Carey}, {Polat}, {Feng}, {Moore}, {VanderPlas},
  {Laxalde}, {Perktold}, {Cimrman}, {Henriksen}, {Quintero}, {Harris},
  {Archibald}, {Ribeiro}, {Pedregosa}, {van Mulbregt}, \& {SciPy 1. 0
  Contributors}}]{SciPy}
{Virtanen}, P., {Gommers}, R., {Oliphant}, T.~E., {et~al.} 2020, Nature
  Methods, 17, 261, \dodoi{10.1038/s41592-019-0686-2}

\bibitem[{Wright(2010)}]{WISE}
Wright, E, e.~a. 2010, The Astronomical Journal, 140, 1868

\bibitem[{{Xie} {et~al.}(2021){Xie}, {Ho}, {Zhuang}, \& {Shangguan}}]{xie2021}
{Xie}, Y., {Ho}, L.~C., {Zhuang}, M.-Y., \& {Shangguan}, J. 2021, \apj, 910,
  124, \dodoi{10.3847/1538-4357/abe404}

\bibitem[{{Yip} {et~al.}(2004{\natexlab{a}}){Yip}, {Connolly}, {Szalay},
  {Budav{\'a}ri}, {SubbaRao}, {Frieman}, {Nichol}, {Hopkins}, {York},
  {Okamura}, {Brinkmann}, {Csabai}, {Thakar}, {Fukugita}, \&
  {Ivezi{\'c}}}]{yip2004a}
{Yip}, C.~W., {Connolly}, A.~J., {Szalay}, A.~S., {et~al.} 2004{\natexlab{a}},
  \aj, 128, 585, \dodoi{10.1086/422429}

\bibitem[{{Yip} {et~al.}(2004{\natexlab{b}}){Yip}, {Connolly}, {Vanden Berk},
  {Ma}, {Frieman}, {SubbaRao}, {Szalay}, {Richards}, {Hall}, {Schneider},
  {Hopkins}, {Trump}, \& {Brinkmann}}]{yip2004b}
{Yip}, C.~W., {Connolly}, A.~J., {Vanden Berk}, D.~E., {et~al.}
  2004{\natexlab{b}}, \aj, 128, 2603, \dodoi{10.1086/425626}

\end{thebibliography}

\end{document}